\documentstyle[11pt,epsf,twoside]{article}
\renewcommand{\thefootnote}{\fnsymbol{footnote}}  

\begin{document}
\setcounter{page}{1}
\noindent
DESY 02--203           \hfill ISSN 0418--9833
\\
hep-ph/0211428        

\vspace{1cm} 
\begin{center}  
{\LARGE \bf 
Status of Electroweak Corrections to Top Pair 
\\[3mm]
Production
\footnote{
Talk presented by T. Riemann at Int. Workshop on Linear Colliders
(LCWS 2002), 26--30 Aug 2002, Jeju Island, Korea
}\footnote{
Work supported in part by the European Community's Human Potential
Programme under contract HPRN--CT--2000--00149 Physics at Colliders
} 
}
  \\ 
\vspace{1.5cm} 
{ 
{\Large J. Fleischer}${}^{1}$\footnote{E-mails:~
fl@physik.uni-bielefeld.de,
leike@theorie.physik.uni-muenchen.de,
\\
\phantom{E-mails:~~~~~~~}
Tord.Riemann@desy.de,
Anja.Werthenbach@desy.de},
~~{\Large A. Leike}${}^{2}$,
~~{\Large T. Riemann}${}^{3}$,
~~{\Large A. Werthenbach}${}^{3,4}$ } 
\\
\vspace{1.cm} 

\small 
{
${}^{1}$~Fakult\"at f. Physik, Universit\"at Bielefeld,  33615
Bielefeld, Germany\\ }
\smallskip
{
${}^{2}$~%
Sektion Physik der Universit\"at M\"unchen, 80333 M\"unchen, Germany
\\ }
\smallskip
{
${}^{3}$~Deutsches Elektronen-Synchrotron, DESY Zeuthen,  15738 Zeuthen, Germany  }  
\\ \smallskip
${}^{4}$~Theory Division, CERN, 1211 Geneva 23, Switzerland

\vspace{1cm}

\end{center}

\vspace{0.2cm} 

\begin{center}
{\Large \bf {Abstract}}
\end{center}
We review the status of electroweak radiative
corrections to top-pair production at a Linear Collider well above the
production threshold. 
We describe the Fortran package {\tt topfit} and
present numerical results at $\sqrt{s}$ = 500 GeV, 1 TeV, and 3 TeV.

\vspace{.5cm}

\setcounter{footnote}{0}  
\renewcommand{\thefootnote}{\arabic{footnote}}  
\setcounter{page}{1}  
\thispagestyle{empty} 

\newpage
\section{Introduction}
Not too much is known about top quarks, and what is known is not as
accurate as desired \cite{Simmons:2002??}.
At a Linear Collider, top-pair production will be one of the dominant
and most interesting processes.  
Very precise measurements are expected.
Therefore, the cross-sections have to be predicted with a precision of
few per mil \cite{Aguilar-Saavedra:2001rg,Abe:2001wn}.
Two quite different experimental set-ups are of interest.
One is the top-pair production threshold region, where one expects to
get precise values of mass and width.
The other one is continuum production at high energy with the hope to
get access to some anomalous behavior,
potentially manifested in abnormal couplings and/or new final state signatures.
Here, we calculate the electroweak Standard Model expectations with
one-loop corrections.
Earlier series of studies are
\cite{Fujimoto:1988hu,Yuasa:1999rg} and
\cite{Beenakker:1991ca,Hollik:1998md}, and a recent one is
\cite{Bardin:2000kn,Andonov:2002xc}.
With our study, detailed comparisons of the diverse results were
undertaken for the first time \cite{Fleischer:2002rn,Fleischer:2002nn}. 
We used the package {\tt DIANA}
\cite{Tentyukov:1999is,Tentyukov:1999yq} for automatic generation of
the diagrams and {\tt FORM} \cite{FORM} for the further symbolic
calculations, and for the numerics
the Fortran packages {\tt FF} \cite{vanOldenborgh:1991yc} and {\tt
  LoopTools} \cite{Hahn:1998yk}.  
\section{The Fortran Package \tt{topfit}}
The package {\tt topfit} \cite{FLRW:2002v0911,FLRW:2002} was written
in order to have a tool for the 
numerical estimation of the electroweak corrections to top-pair
production. 
We wanted also to have some flexibility for an easy comparison with
other codes.
As a result, the user of our program may  switch on and
off several options and may adjust input parameters.
The list in Table \ref{table-1} is by far not complete.
Of course, the usual standard model parameters in the on-mass-shell
scheme (particle masses and $\alpha_{em}(0)$) may be chosen.
The numerical input is as in \cite{Fleischer:2002rn}.
\begin{table}[bht]
\begin{center}
\begin{tabular}{|c|c|}
\hline 
Flag & Description
\\ \hline 
\hline 
{\tt IFINAL} & choice of final state particle 
\\ \hline 
{\tt IQED} & inclusion of photonic corrections
 \\ \hline 
{\tt CNINI} & initial state corrections
 \\ \hline 
{\tt CNFIN} &  final state corrections
\\ \hline 
{\tt CNINT} & interference corrections
\\ \hline 
{\tt IWEAK} & Born or use {\tt LoopTools} or use {\tt FF}
 \\ \hline 
{\tt GAMS} & choice of $Z$ width
 \\ \hline 
{\tt IQEDAA} & running of $\alpha_{em}$
 \\ \hline 
{\tt IPHOTM} & finite photon mass or dimensional regularisation of
the IR divergency 
 \\ \hline 
{\tt IDCOST} & top quark angular distribution 
 \\ \hline 
{\tt ICOSTI} & integrated cross-section and asymmetry
 \\ \hline 
{\tt IHARD} & inclusion of hard bremsstrahlung
 \\ \hline 
{\tt SPRCUT} & photonic phase space cut
\\ 
\hline
\end{tabular}
\end{center}
\caption[]{\label{table-1}
A collection of flags of {\tt topfit}; more details may be found in
\cite{FLRW:2002v0911}. 
}
\end{table}
Three different options may be chosen for the output:
\begin{itemize}
\item Differential and integrated cross-sections and asymmetries in
  the Standard Model  
\item Cross-sections and asymmetries with photonic corrections only,
  at fixed Born couplings 
\item Six weak form factors, with/without running $\alpha_{em}$
\end{itemize} 
The latter output might be useful for a Monte Carlo approach to QED
and QCD corrections, but also for an estimate of the weak corrections.
\section{Numerical Results and Conclusions}
\begin{figure}[thb]
 \centerline{\epsfysize=15cm\epsffile{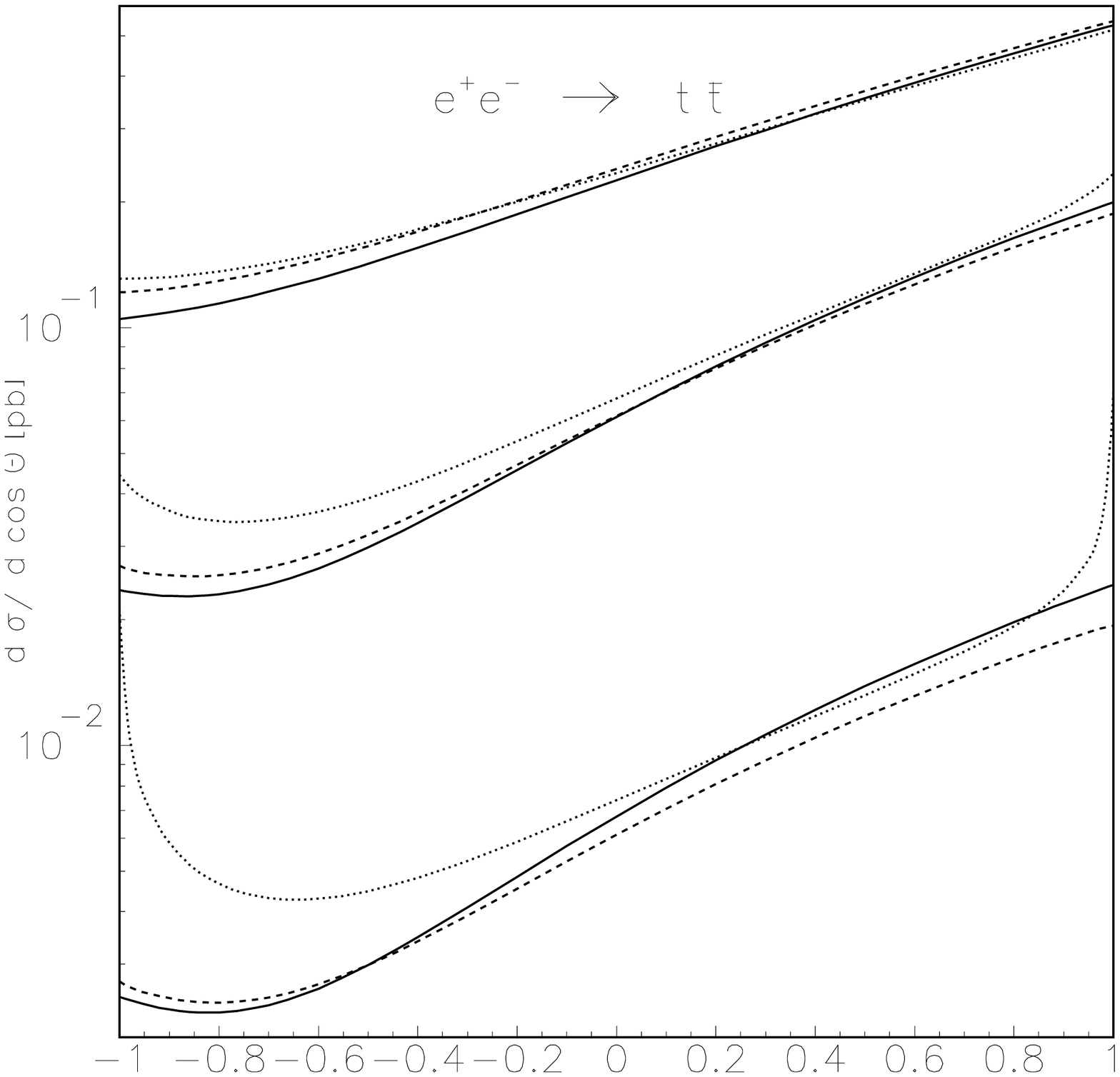}}
\vspace*{-0.7cm}
\caption{\label{eecc}
{\it 
Differential cross-sections in Born approximation (solid lines), with
weak corrections (dashed), and with full electroweak corrections (no
cut; dotted). From above: $\sqrt{s}$ = 0.5, 1, 3 (in TeV). 
}}
\end{figure}

For comparisons with the results of other groups, with very
satisfactory numerical agreements, we refer
to \cite{Fleischer:2002rn,Fleischer:2002nn}.
Of course, one has to bear in mind that these comparisons control not
more than what is called `technical precision'.
In Figure \ref{eecc} we show the order of magnitude of the
cross-sections, and in Table \ref{table-2} purely weak form factors at
$\sqrt{s}$ = 3 TeV.
To define the normalizations of $F_3^{11}$ and $F_3^{51}$, we
mention that 
$d\sigma \sim  Re [ 
(u^2 + t^2 + 2m_t^2s) |F_1^{11}|^2 + 2m_t (ut - m_t^4) 
(F_1^{11*}F_3^{11} + F_1^{51*}F_3^{51})]$ + \ldots, and 
$t = s(1-\beta_t\cos\theta)/2$.

\begin{table}[bht]
\begin{center}
\begin{tabular}{|c|c|c|}
\hline 
f.f. & Born & Born + weak corrections 
\\ \hline \hline 
 $F_1^{11}$ & --3.4822175~E--09,~0&--2.4672033~E--09,~+5.6471323~E--12
\\
 $F_1^{15}$ & +2.0992410~E--10,~0&--4.6533609~E--10,~--3.4235887~E--10
\\
 $F_1^{51}$ & +7.5582979~E--10,~0 &+1.4831421~E--10,~--2.6754148~E--10
\\
 $F_1^{55}$ & --1.8476412~E--09,~0&--1.4913239~E--09,~+3.4972393~E--10
\\
 $m_tF_3^{11}$ & 0,~0             &+2.9895163~E--12,~--6.6708986~E--13
\\
 $m_tF_3^{51}$ & 0,~0             &--2.4939160~E--12,~+9.1292861~E--13
\\
\hline
\end{tabular}
\end{center}
\caption[]{\label{table-2}\it
Weak form factors for $d\sigma/d\cos\theta$ 
at $\sqrt{s}$=3 TeV, $\cos\theta$=0.7. 
They yield 
$\sigma_B$=0.076266014 pb and 
$\sigma_{weak}$=0.012482585 pb, correspondingly.
The 
normalization corresponds to $F_{1,Born}^{11,{\gamma}} = e^2Q_eQ_t/s$ (see also
\cite{Fleischer:2002rn}); some flags chosen: {\tt IWEAK} = 1, {\tt
  GAMS} = --1, {\tt IQED}
= {\tt IQEDAA} = 0. 
}
\end{table}

The treatment of the one-loop electroweak corrections to top-pair production is
well under control up to $\sqrt{s}$ = 3 TeV.
Of course, there is much Standard Model physics to be included in
addition: higher 
orders, notably in the photonic part, but also numerically large QCD
corrections, and finally also phenomena like top-quark decays, other
background reactions (of different signatures like $e^+e^-\to tWb,
tbl_1l_2$, $6f$), or beamstrahlung. 

\vspace*{-1mm}
{\bf Acknowledgments.}
~T.R. and A.W. would like to thank the organizers of LCWS 2002 for the
opportunity to visit an interesting,
well-organized, and pleasant conference.
F.J. and A.L. would like to thank DESY for support.

\providecommand{\href}[2]{#2}

\end{document}